\documentclass{article}
\usepackage{latexsym}
\usepackage{graphics}
\usepackage{color}
\usepackage{amsmath, amsthm, amssymb}
\newtheorem{proposition}{Proposition}

\newtheorem{theorem}{Theorem}
\newtheorem{lemma}{Lemma}
\newtheorem{corollary}{Corollary}

\def\tr{{\rm tr}}
\title{An extension principle for the Einstein-Vlasov system in spherical 
symmetry}
\author{Mihalis Dafermos\thanks{University of Cambridge,
Department of Pure Mathematics
and Mathematical Statistics, Wilberforce Road, Cambridge CB3 0WB, United 
Kingdom}
\and Alan D.~Rendall\thanks{Max Planck Institute for
Gravitational Physics, Albert Einstein Institute, Am Muehlenberg 1,
14476 Golm, Germany}}
\begin{document}
\maketitle
\begin{abstract}
We prove that ``first singularities'' in 
the non-trapped region of the maximal development of spherically
symmetric asymptotically flat data for the Einstein-Vlasov
system must necessarily emanate from the center. 
The notion of ``first'' depends only on the causal
structure and can be described in the language of terminal
indecomposable pasts (TIPs).
This result suggests a local approach to
proving weak cosmic censorship for this system. It can also be used 
to give the first proof of the formation of black holes by the
collapse of collisionless matter from regular initial configurations. 
\end{abstract}

\section{Introduction}
A fundamental problem in mathematical relativity is
to resolve the so-called \emph{weak cosmic censorship conjecture}, the 
statement
that for ``reasonable'' Einstein-matter systems, generic asymptotically
flat data do
not lead to singularities visible from infinity. 

The notion of ``reasonable'' above is of course not a precise one,
and depends very much on the context one has in mind. A natural 
matter source for models is provided by kinetic theory. The simplest
example is then a self-gravitating collisionless gas. The study of 
the equations describing such a gas,
the Einstein-Vlasov system, was initiated
by Choquet-Bruhat in~\cite{ycb}, where the existence of a unique
maximal development was proven for the Cauchy problem.

The problem of weak cosmic censorship concerns the global behaviour of
the maximal development for asymptotically flat initial data.
Given the current state of the art in nonlinear evolution equations,
symmetry must be imposed on initial data for there to be any hope of making 
progress. The global study of the initial value problem for
the Einstein-Vlasov equations for spherically symmetric
asymptotically flat initial data was begun in~\cite{rr}, 
where, in particular, it was proven that for sufficiently small 
initial data, the maximal development was future causally
geodesically complete. The analysis took place in so-called Schwarzschild 
coordinates.
In~\cite{rrs}, an extension principle was proven, again in these coordinates,
saying in particular that if the solution stopped existing after finite 
coordinate time $t$, there was necessarily a singularity at the center.
These results were meant to provide a first
step for a \emph{global} existence theorem in Schwarzschild
coordinates. If this coordinate system could then
be shown to cover the domain of outer communications, 
and if null infinity could moreover be shown to be 
complete, this would then imply a proof of weak cosmic
censorship for this system.

There is another approach to the problem of weak cosmic censorship,
due to  Christodoulou~\cite{chr:ins}, for the problem
of a self-gravitating spherically symmetric scalar field. 
Christodoulou showed that initial data leading
to a naked singularity was codimension $1$ in the space of all initial data.
This was shown by embedding such exceptional data in a one-dimensional subset
of the space of initial data, such that all other initial data 
in this subset evolved to a spacetime with the following property,
which can be expressed in the language of causal sets \cite{gkp}.
Given a terminal indecomposable past (TIP) with compact intersection with 
the Cauchy surface, then the domain of dependence of any open set containing 
this intersection contains a trapped surface.
The statement that this latter property is true
for generic initial data can be termed the \emph{trapped surface conjecture}.
{}From this property, the completeness of null infinity was then inferred,
proving weak cosmic censorship.

It turns out that the relation between the existence of trapped
surfaces and the completeness of null infinity is quite general.
Specifically, in~\cite{md}, it was proven that a weaker version
of the trapped surface conjecture is sufficient to prove weak
cosmic censorship for a wide variety of matter in spherical
symmetry. In particular, the completeness of null infinity
follows from the existence of a \emph{single} trapped or marginally
trapped surface in the maximal 
development. The only really restrictive 
hypothesis on the matter is
that ``first'' singularities necessarily emanate from the center.
Here, the notion of ``first'' is tied to the causal structure and
can be formulated in terms of TIPs.

The goal of this paper is to prove that the above mentioned hypothesis 
of~\cite{md} is indeed satisfied by the Einstein-Vlasov system.
As noted before, extension principles similar in spirit to this
one have been proven before (cf.~\cite{rrs,adr2}). 
These earlier results,
however, concern the portion of the development
of the Einstein-Vlasov system
covered by particular coordinate systems. 
Thus, these previous results,
as far as they concern the maximal development itself,
are weaker than the results presented here, 
and in particular, are not sufficient
to deduce the assumptions of~\cite{md}.\footnote{Of course,
the results of~\cite{rrs, adr2} 
also say something about the behaviour 
of the coordinate system
to which they apply, something not addressed here.}

Finally, we make the following remark:
In view of~\cite{adr1}, there do exist
spherically symmetric asymptotically flat initial
data for the Einstein-Vlasov system possessing a trapped surface.
Thus, the results of this paper provide in particular the first
proof of the existence of solutions for collisionless matter
representing the formation of a black hole.

\section{Initial data}
Initial data in this paper are always given as follows:
\begin{enumerate}
\item
We have a $C^\infty$ Riemannian manifold $(\Sigma, \bar{g})$,
together with an additional symmetric $2$-tensor $K_{ab}$,
such that there do not exist closed antitrapped surfaces in the data,
and a compactly supported function $f_0$ defined on the tangent bundle of 
$\Sigma$, such that these satisfy
\begin{eqnarray}
\bar R-K_{ab}K^{ab}+(\tr K)^2&=&16\pi\int f_0(p^a)p^ap_a/(1+p^ap_a)^{1/2}
\sqrt{\bar g}dp^1dp^2dp^3  \nonumber\\
\nabla_aK^a{}_b-\nabla_b(\tr K)&=&8\pi\int f_0(p^a)p_a\sqrt{\bar g}
dp^1dp^2dp^3 \nonumber 
\end{eqnarray}
Here the metric $\bar g$
is used to move indices and to define the trace and covariant derivative.
$\bar R$ is the scalar curvature of $\bar g$ and $\sqrt{\bar g}$ the 
square root of its determinant.
\item
A smooth $SO(3)$ action on $\Sigma$ such that
$\bar{g}$, $K_{ab}$, $f_0$ are preserved, and such that 
$\Sigma/SO(3)$ inherits naturally the structure of a $1$-dimensional manifold.
\end{enumerate}

Here and throughout this paper physical units are chosen so that the 
gravitational constant has the numerical value unity. We recall the 
definition of a closed antitrapped surface. Let $S$ be
a surface in $\Sigma$ which is closed, i.e. compact without boundary.
Suppose that there is a preferred choice $n^a$ of an outward normal to
this surface and let $\sigma_{ab}$ be the second fundamental form
of $S$ in $\Sigma$ corresponding to the outward normal. Then $S$
is said to be antitrapped if $\tr\sigma<-\tr K+K_{ab}n^an^b$.

\section{The maximal development}
The theorem of Choquet-Bruhat~\cite{ycb}, applied
to the data considered here,
together with a standard argument on preservation of symmetry,
yields
\begin{proposition}
There exists a unique $C^\infty$ collection $(\mathcal{M},g, f)$
such that
\begin{enumerate}
\item
$g$ and $f$ satisfy the Einstein-Vlasov equations
\item
($\mathcal{M}$, $g$) is globally hyperbolic,
\item
$(\mathcal{M},g,f)$ induces the initial data $(\Sigma,\bar{g},K,f_0)$ and 
$\Sigma$ is a Cauchy surface
\item
Any other collection $(\mathcal{M}, g, f)$ with these properties 1-3
can be embedded in the given one.
\end{enumerate}
Moreover, $SO(3)$ acts smoothly by isometry on $\mathcal{M}$ and preserves
$f$, and 
$\mathcal{Q}=\mathcal{M}/SO(3)$
inherits the structure of
a time-oriented $2$-dimensional Lorentzian manifold,
with timelike boundary $\Gamma$, the center. 

\end{proposition}

Let $\pi:\mathcal{M}\to\mathcal{Q}$ denote
the natural projection.
On $\mathcal{Q}$ we can define the so-called
area-radius function
\[
r(p)=\sqrt{{\rm Area}(\pi^{-1}(p))/4\pi}
\]
We have $r(p)=0$ iff $p\in\Gamma$.
We can always
choose global future directed null coordinates
on $\mathcal{Q}$, i.e.~such that the metric
takes the from $-\Omega^2dudv$.
The metric of $\mathcal{M}$ then takes the form:
\begin{equation}\label{metric}
-\Omega^2dudv+r^2\gamma
\end{equation}
where $\gamma=\gamma_{AB}dx^Adx^B$ is the standard metric on $S^2$
and $x^A$, $A=2,3$, are local coordinates on $S^2$.
Let $u$ and $v$ be chosen so that $\frac\partial{\partial u}$
points ``inwards'' and $\frac\partial{\partial v}$ ``outwards''.
Such definitions are meaningful in view of the assumption
of asymptotic flatness. 
We define
\[
\nu=\partial_u r
\]
\[
\lambda=\partial_v r
\]
The assumption of no antitrapped surfaces initially means by definition
that
\begin{equation}
\label{prosnmo}
\nu<0
\end{equation}
holds on the initial hypersurface. It follows that it holds throughout 
$\mathcal{Q}$ as a consequence of the Einstein equations and the
dominant energy condition \cite{chr}.

We shall call the region where $\lambda>0$,
the \emph{regular region}, and denote it $\mathcal{R}$.
We call the region where $\lambda=0$ the \emph{marginally
trapped region}, and denote it by $\mathcal{A}$, and finally,
we shall can the region where $\lambda<0$ the \emph{trapped region}, 
and denote it by $\mathcal{T}$.

\section{The extension theorem}
The extension principle proven in this paper will apply to
a region $\mathcal{D}\subset \mathcal{Q}$ with Penrose diagram:
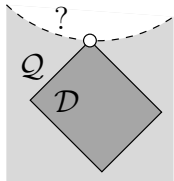
\[
\begin{picture}(0,0)%
\includegraphics{figure.pstex}%
\end{picture}%
\setlength{\unitlength}{3947sp}%
\begingroup\makeatletter\ifx\SetFigFont\undefined%
\gdef\SetFigFont#1#2#3#4#5{%
  \reset@font\fontsize{#1}{#2pt}%
  \fontfamily{#3}\fontseries{#4}\fontshape{#5}%
  \selectfont}%
\fi\endgroup%
\begin{picture}(1066,1134)(4268,-4112)
\put(4576,-3661){\makebox(0,0)[lb]{\smash{\SetFigFont{12}{14.4}{\rmdefault}{\mddefault}{\updefault}{\color[rgb]{0,0,0}$\mathcal{D}$}%
}}}
\put(4351,-3436){\makebox(0,0)[lb]{\smash{\SetFigFont{12}{14.4}{\rmdefault}{\mddefault}{\updefault}{\color[rgb]{0,0,0}$\mathcal{Q}$}%
}}}
\put(4576,-3136){\makebox(0,0)[lb]{\smash{\SetFigFont{12}{14.4}{\rmdefault}{\mddefault}{\updefault}{\color[rgb]{0,0,0}$?$}%
}}}
\end{picture}

\]
\noindent
(i.e.~a subset $\mathcal{D}=[u_1,u_2]\times[v_1,v_2]\setminus
(u_2,v_2)$)
such that 
\[
\mathcal{D}\subset\mathcal{R}\cup\mathcal{A}.
\]
Let $\mathcal{C}_{{\rm in}}$ and $\mathcal{C}_{{\rm out}}$ be the parts of the 
boundary 
of $\mathcal{D}$ defined by $v=v_1$ and $u=u_1$ respectively.
One can think of $\mathcal{D}$ as the ``top'' of a non-trapped
non-central indecomposable past (IP) corresponding to a candidate ``first'' 
singularity. In this language, the result of this paper is that such an IP 
cannot be a TIP, i.e.
\begin{theorem}
\label{extension}
If $\mathcal{D}\subset\mathcal{Q}$, then $\mathcal{D}\subset
J^-(q)$ for a $q\in\mathcal{Q}$.
\end{theorem}
The theorem thus says that there is no singularity of this form after all!

As one might expect,
the proof of Theorem~\ref{extension} proceeds by obtaining \emph{a priori}
estimates in $\mathcal{D}$ and then applying an appropriate local existence 
result.
The \emph{a priori} 
estimates make use of a certain energy flux along null 
hypersurfaces. This fact, together with the fact that regular null coordinates
can always be chosen, makes it natural to stick to these.
We give the form of the equations in local null coordinates in
the next two sections. Then, in Section~\ref{locsec}, we formulate a local 
existence theorem (Proposition~\ref{local}) 
for a double characteristic initial value problem. The ``time'' of
existence, in the sense of null coordinates, will depend only
on the $C^2$ norm of the metric and the $C^1$ norm (and the support)
of $f$. 
We obtain energy estimates in
Section~\ref{enestsec}, and use these, together 
with the structure of the Vlasov equation, 
to derive in Sections~\ref{c1sec}--\ref{c2sec} \emph{a priori}
estimates for the norm of Proposition~\ref{local}. The proof
of Theorem \ref{extension} will follow immediately in Section~\ref{theproof}.
Finally, in Section~\ref{appli},
we state two applications of our results, discussed already in the
Introduction.

The above theorem depends on having a well-behaved matter model and the
analogous result must be expected to fail for dust. This is illustrated by 
the Penrose diagram Fig. 1 in \cite{yodzis}.

\section{The Einstein equations in null coordinates}
The reader should consult \cite{chr} for general facts about
the initial value problem in spherical symmetry. When specialized
to this case, the Einstein equations are:
\begin{equation}
\label{evol1}
\partial_u\partial_vr =-\frac{\Omega^2}{4r}-\frac{1}r\lambda\nu
+4\pi rT_{uv},
\end{equation}
\begin{equation}
\label{evol2}
\partial_u\partial_v\log \Omega=
-4\pi T_{uv}+\frac{\Omega^2}{4r^2}+\frac{1}{r^2}\lambda\nu
-\frac{\pi\Omega^2}{r^2}\gamma^{AB}T_{AB},
\end{equation}
\begin{equation}
\label{constr1}
\partial_v(\Omega^{-2}\partial_v r)=-4\pi rT_{vv}\Omega^{-2},
\end{equation}
\begin{equation}
\label{constr2}
\partial_u(\Omega^{-2}\partial_u r)=-4\pi rT_{uu}\Omega^{-2}.
\end{equation}
The former two equations can be viewed as wave equations for $r$ and $\Omega$,
while the latter two equations can be viewed as constraint equations 
on null hypersurfaces. A specific choice of matter model, such as a
collisionless gas, leads to expressions for the components of
the energy-momentum tensor.

\section{The Vlasov equation}
To describe the Vlasov equation in local coordinates, we need a coordinate 
system on $T\mathcal{M}$. Let $p^u$, $p^v$, and $p^A$ denote the functions
on $T\mathcal{M}$, defined by writing an arbitrary
${\bf X}\in T\mathcal{M}$ as
\[
{\bf X}=p^u\frac\partial{\partial u}+p^v
\frac\partial{\partial v}
+p^A\frac\partial{\partial x^A}
\]
Together with the pull-back of the coordinates on spacetime these functions
define a local coordinate system on $T\mathcal{M}$.

Let $P\subset T\mathcal{M}$ be defined
by 
\[
P=\{g({\bf X},{\bf X})=-1\}, 
\]
where
${\bf X}$ ranges over future-pointing vectors.
We call $P$ the mass shell. It follows that
\begin{equation}\label{massshell}
-\Omega^2 p^up^v+r^2\gamma_{AB}p^Ap^B=-1
\end{equation}
We use $p^u$, $p^A$ and the pull-back of the coordinates on spacetime
to define coordinates on $P$ and $p^v$ is regarded as a function of 
these coordinates defined by the relation (\ref{massshell}).  
The Vlasov equation is an equation for
a non-negative function
\[
f:P\to{\bf R}
\]
which, in the case that $f$ is spherically symmetric, is given by
\begin{eqnarray}
\label{vlasoveq}
\nonumber
p^u\frac{\partial f}{\partial u}+p^v
\frac{\partial f}{\partial v}
&=&(\partial_u(\log\Omega^2)(p^u)^2+2\Omega^{-2}
r\lambda\gamma_{AB}p^Ap^B)\frac{\partial f}{\partial
 p^u}\\
&&\hbox{}+2r^{-1}(\nu p^u+\lambda p^v)p^A
\frac{\partial f}{\partial p^A}.
\end{eqnarray}
In deriving this we have used the expressions for the Christoffel symbols
given in Appendix A and the fact that a spherically symmetric
function $f$ on the mass shell is a function of the variables
$u$, $v$, $p^u$, and $\gamma_{AB} p^A p^B$. This implies the identity
\[ 
p^A\frac{\partial f}{\partial x^A}=\Gamma^A_{BC}p^Bp^C\frac{\partial f}
{\partial p^A}
\]
which has been used to simplify the Vlasov equation. Note that both
the expressions $\gamma_{AB}p^Ap^B$ and $p^A\frac\partial{\partial p^A}$
have a meaning independent of the particular choice of coordinates $x^A$ 
on $S^2$.


Finally, to close the system, we must define the
energy-momentum tensor. We first note that 
for any point $q\in\mathcal{M}$, it follows that $P_q$,
as a spacelike hypersurface in $T_q\mathcal{M}$,
inherits a volume form from the Lorentzian metric.
In local coordinates
this volume form can be written $r^2(p^u)^{-1}dp^u\sqrt{\gamma}dp^Adp^B$ 
or alternatively $r^2(p^v)^{-1}dp^v\sqrt{\gamma}dp^Adp^B$, where 
$\sqrt{\gamma}$ is the square root of the determinant of $\gamma_{AB}$.
We then have
\begin{equation}
\label{asteraki}
T_{ab}=
\int_0^\infty\int_{-\infty}^{\infty}
\int_{-\infty}^{\infty}
r^2p_ap_bf(p^u)^{-1}\sqrt{\gamma}dp^udp^Adp^B,
\end{equation}
where $p_a=g_{ab}p^b$.
It follows immediately that this matter model
satisfies the energy conditions:
\begin{equation}
\label{suv9nkn-evergeia}
T_{uv}\ge0, T_{vv}\ge0, T_{uu}\ge0
\end{equation}

\section{A local existence theorem}
\label{locsec}

To prove our extension theorem, we will certainly need to appeal to some
sort of local existence theorem. In particular, it is the norm in this
theorem that will tell us what quantities we must 
bound \emph{a priori} in $\mathcal{D}$.
In principle, one could try to prove estimates so as to apply 
the local existence result of~\cite{ycb}.
For various reasons, however, the following
local existence theorem for a characteristic initial 
value problem will be more convenient:

\begin{proposition}
\label{local}
Let $k\ge2$.
Let $\Omega, r$ be positive $C^k$-functions
defined on $[0,d]\times \{0\}\cup \{0\}\times[0,d]$,
and let $f$ be a non-negative $C^{k-1}$ function defined on the 
part of the mass shell over $[0,d]\times\{0\}\cup\{0\}\times[0,d]$.
Suppose that equations $(\ref{constr1})$, $(\ref{constr2})$ hold on
$\{0\}\times[0,d]$ and $[0,d]\times \{0\}$ respectively, where $T_{uu}$
and $T_{vv}$ are defined by $(\ref{asteraki})$, and 
suppose in addition that the $C^k$ compatibility condition holds at $(0,0)$.
Define the norm:
\begin{eqnarray*}
N_u&=&\sup_{[0,d]\times \{0\}}
\{|\Omega|, |\Omega^{-1}|, |\partial_u \Omega|, |\partial_u^2\Omega|,
|r|, |r|^{-1}, |\partial_u r|, |\partial_u^2 r|,\\
&&\hbox{} S, |f|,
|\partial_u f|, |\partial_{p^u}f|, |\partial_{p^A}f|_\gamma \},
\end{eqnarray*}
\begin{eqnarray*}
N_v&=&\sup_{\{0\}\times[0,d]}
\{|\Omega|,|\Omega^{-1}|, |\partial_v\Omega|, |\partial_v^2 \Omega|,
|r|, |r|^{-1}, |\partial_v r|, |\partial_v^2 r|,\\
&&\hbox{}S, |f|,
|\partial_v f|, |\partial_{p^u}f|, |\partial_{p^A}f|_\gamma \},
\end{eqnarray*}
\[
N=\sup \{N_u, N_v\},
\]
were $S$ denotes the supremum of $(p^u)^2+(p^v)^2+\gamma_{AB}p^Ap^B$
on the support of $f$ and $|v_A|_\gamma=(\gamma^{AB}v_Av_B)^{1/2}$.
Then there exists a $\delta$,
depending only on $N$,
and $C^{k}$ functions (unique among $C^2$ functions)
$r, \Omega$ and a $C^{k-1}$ function (unique among $C^1$ functions)
$f$,
satisfying equations $(\ref{evol1})$, $(\ref{evol2})$,
$(\ref{constr1})$, $(\ref{constr2})$, $(\ref{vlasoveq})$  in 
$[0,\delta^*]\times[0,\delta^*]$, where
$\delta^*=\min\{d,\delta\}$, such that
the restriction of these functions
to $[0,d]\times\{0\}\cup\{0\}\times[0,d]$
is as prescribed.
\end{proposition}
\emph{Proof}. See Appendix~\ref{localproof}.

The compatibility conditions referred to in the statement of the 
proposition are as follows. The data includes the values of the function
$f$ on the part of the mass shell over $[0,d]\times \{0\}$. All 
derivatives of $f$ tangential to this manifold can be calculated by
direct differentiation. By using the field equations transverse 
derivatives (and thus all derivatives) of $f$ can be computed up to
order $k-1$. In a similar way, all derivatives up to order $k-1$
can be computed on $\{0\}\times [0,d]$. The condition that derivatives
determined in these two different ways agree at $(0,0)$ is what is referred to
above as the $C^k$ compatibility condition.

Let us add the remark that, defining $g$ on $\mathcal{M}$
by (\ref{metric}), the above gives rise to a solution of
the Einstein-Vlasov equations
upstairs, with the obvious relation to characteristic
data, interpreted upstairs.

\section{Energy estimates}
\label{enestsec}

A fundamental fact about the analysis of spherically
symmetric Einstein matter systems in the non-trapped region
is the existence of energy estimates.

To describe these, let us first settle
for a particular null-coordinate description
of the set $\mathcal{D}$.
We normalize our $u$-coordinate such that $\nu=-1$ along
$\mathcal{C}_{{\rm in}}$.
For the $v$ coordinate, we first define the quantity
\[
\kappa=-\frac14\Omega^{2}\nu^{-1}.
\]
and then define $v$ such that $\kappa=1$ along
$\mathcal{C}_{{\rm out}}$. 
$\mathcal{D}$ is thus given
by $[0,U]\times [0,V]\setminus\{ (U,V)\}$.

The concept of energy in spherical symmetry is given by
the so-called \emph{Hawking mass}, given by:
\[
m=\frac r2(1-\partial^ar\partial_ar)=
\frac r2(1-2g^{uv}\partial_u r\partial_v r)
=\frac r2(1+4\Omega^{-2}\lambda\nu).
\]
We will also introduce the so-called \emph{mass-aspect} function
\[
\mu=\frac{2m}r.
\]
Note that
\begin{equation}
\label{lambda=}
\kappa(1-\mu)=\lambda.
\end{equation}
{}From $(\ref{evol1})$--$(\ref{constr2})$, we compute the identities:
\begin{eqnarray}
\label{pum}
\nonumber
\partial_um     &=&8\pi   r^2\Omega^{-2}(T_{uv}\nu-T_{uu}\lambda)\\
                &=&-2\pi\kappa^{-1}r^2T_{uv}+2\pi\frac{1-\mu}{\nu}r^2T_{uu}
\end{eqnarray}
\begin{eqnarray}
\label{pvm}
\nonumber
\partial_vm     &=&     8\pi r^2\Omega^{-2}(T_{uv}\lambda-T_{vv}\nu)\\
                &=&     -2\pi\frac{1-\mu}{\nu}r^2T_{uv}+
                        2\pi\kappa^{-1}r^2T_{vv}.
\end{eqnarray}

The first point to note is that the signs of $(\ref{pum})$ and $(\ref{pvm})$,
together with the signs of $\lambda$ and $\nu$, give \emph{a priori} bounds 
for both $r$ and $m$.
Indeed, set
\[
m_0=m(U,0)\ge0,
\]
\[
r_0=r(U,0)>0,
\]
\[
M=m(0,V),
\]
\[
R=r(0,V).
\]
By $(\ref{prosnmo})$ and the fact
that $\mathcal{D}\subset\mathcal{R}\cup\mathcal{A}$,
we have
that
\begin{equation}
\label{r}
r_0\le r\le R
\end{equation}
throughout $\mathcal{D}$.
On the other hand, $(\ref{pum})$, $(\ref{pvm})$ and
$(\ref{suv9nkn-evergeia})$
give $\partial_um\le0$, $\partial_vm\ge0$,
and thus
\begin{equation}
\label{maza}
m_0\le m\le M.
\end{equation}

Now we make a trivial observation. In view
of the fact that we have the \emph{a priori} bounds
$(\ref{maza})$, if we reexamine the equations
$(\ref{pum})$, $(\ref{pvm})$, keeping in mind 
that both terms on the right hand side have the same sign,
we obtain the bounds:
\begin{equation}
\label{EE1}
\int_{v_1}^{v_2}\frac{2\pi(1-\mu)}{-\nu}r^2 T_{uv}(u,v)dv\le M-m_0,
\end{equation}
\begin{equation}
\label{EE2}
\int_{v_1}^{v_2}2\pi\kappa^{-1}r^2T_{vv}(u,v)dv\le M-m_0,
\end{equation}
\begin{equation}
\label{EE3}
\int_{u_1}^{u_2}2\pi\kappa^{-1}r^2T_{uv}(u,v)du\le M-m_0,
\end{equation}
\begin{equation}
\label{EE4}
\int_{u_1}^{u_2}\frac{2\pi(1-\mu)}{-\nu}r^2T_{uu}(u,v)du\le M-m_0.
\end{equation}
These will be our \emph{energy estimates}. 

As we shall see, our use of the above estimates will not quite
be symmetric for $u$ and $v$. The reason is this:
The ``constraint'' equation  $(\ref{constr2})$
can be seen to be equivalent to the
following equation for $\kappa$:
\begin{equation}
\label{puk}
\partial_u\kappa=4\pi r\nu^{-1}T_{uu}\kappa.
\end{equation}
{}From $(\ref{prosnmo})$, $(\ref{puk})$ and $(\ref{suv9nkn-evergeia})$, 
we see
immediately
\begin{equation}
\label{kappa}
0<\kappa\le1
\end{equation}
throughout $\mathcal{D}$, i.e.~$\kappa^{-1}\ge 1$. This means that
\emph{a priori} we control $\int T_{vv}dv$,
but not $\int T_{uu}du$.

Finally, note that
we can rewrite equation $(\ref{evol1})$
as
\begin{equation}
\label{vu-e3is}
\partial_v\nu=2r^{-2}\kappa\nu m+ 4\pi rT_{uv},
\end{equation}
or alternatively
\begin{equation}
\label{lambda-e3is}
\partial_u\lambda=2r^{-2}\kappa\nu m+4\pi rT_{uv}.
\end{equation}
Thus, integrating $(\ref{vu-e3is})$, 
in view of $(\ref{kappa})$, $(\ref{maza})$,
$(\ref{r})$, and $(\ref{suv9nkn-evergeia})$,
we have that
\begin{equation}
\label{nubound}
\nu\ge -e^{2r_0^{-2}MV}=-\tilde{N}.
\end{equation}

\section{$C^1$ estimates for the metric}
\label{c1sec}
So far, we have not used the Vlasov equation, only the energy
condition $(\ref{suv9nkn-evergeia})$. Indeed, all estimates
obtained so far are familiar from the results of~\cite{md}.
To go further, we must use the 
Vlasov equation itself and the special structure 
of the energy-momentum tensor. In this section, we shall
estimate the support of $f$ and show $C^1$ estimates for the metric.

Before proceeding, let us give names
to bounds on certain quantities on the initial
segments $\mathcal{C}_{\rm in}\cup\mathcal{C}_{\rm out}$.
Define
\[
G=
\max\left\{\sup_{[0,U]\times\{0\}}|\partial_u\log \Omega^2|,
\sup_{\{0\}\times[0,V]}|\partial_v\log\Omega^2|\right\}
\]
\[
F=\sup_{\pi_1^{-1}(\{0\}\times[0,V]\cup[0,U]\times\{0\})}
f,
\]
where $\pi_1$ denotes the projection from the mass shell,
define $\Sigma$ to be supremum of the radius
of support of $f$ in the $p^v$ and $p^u$ directions
along $\pi_1^{-1}(\{0\}\times[0,V]
\cup[0,U]\times\{0\})$, and define
$X$ be the supremum of $\gamma_{AB}p^Ap^B$ over the support of $f$.

Let us note first two easy bounds.
Clearly,
\[
0\le f\le F
\]
throughout the mass shell over $\mathcal{D}$.
Moreover, by $(\ref{r})$ and conservation of
angular momentum applied to geodesics, it follows that
\begin{equation}
\label{am}
\gamma_{AB}p^Ap^B(x)\le X
\end{equation}
for any $x\in P$ in the support of $f$ over $\mathcal{D}$. 
In particular, in the expressions defining energy-momentum,
we can thus always replace 
an integral over the variables $p^A$ by the integral over the
ball of radius $X$ about the origin.

We have the following:
\begin{lemma}
The inequality
\[
-g_{uv}g^{AB}T_{AB}\le 2T_{uv}
\]
holds throughout $\mathcal{D}$.
\end{lemma}

\noindent\emph{Proof.} 
The inequality is equivalent to the statement that the trace of the 
energy-momentum tensor is non-positive. This holds for collisionless
matter independently of symmetry assumptions. It is proved 
straightforwardly by taking a trace in the formula defining the 
energy-momentum tensor in general coordinates with the spacetime metric.
$\Box$


We can rewrite $(\ref{evol2})$ as
\begin{eqnarray}
\label{kampulotnta}
\partial_u(\partial_v \log \Omega^2)&=&
-8\pi T_{uv}-4\kappa m r^{-3}\nu+8\pi \kappa\nu r^{-2}\gamma^{AB}T_{AB}.
\end{eqnarray}
Integrating  $(\ref{kampulotnta})$,
applying the above lemma, the energy estimate $(\ref{EE3})$,
and the bounds $(\ref{r})$, $(\ref{maza})$, $(\ref{kappa})$,
we estimate $\partial_v\log \Omega^2$:
\begin{eqnarray}
\label{G'}
\nonumber
|\partial_v\log\Omega^2|        &\le&   G+\left|\int{8\pi T_{uv}du}
			-\int{8\pi\kappa\nu r^{-2}\gamma^{AB}T_{AB}du}\right|
                -\int{4\kappa mr^{-3}\nu du}
                \\
\nonumber
                &\le&   G+\int{4\kappa r^{-2}
                        \left(2\pi r^2\kappa^{-1}T_{uv}\right)du}
                -\int{4\kappa mr^{-3}\nu du}\\
\nonumber
                &\le&   G+4r_0^{-2}\int{
                        2\pi r^2\kappa^{-1}T_{uv}du}
                -\int{4\kappa mr^{-3}\nu du}\\
\nonumber
                &\le&   G+4r_0^{-2}\int{
                        2\pi r^2\kappa^{-1}T_{uv}du}
                -\int{4\kappa mr^{-3}\nu du}\\
\nonumber
        &\le&   G+4r_0^{-2}(M-m_0)+2(r_0^{-2}-R^{-2})M\\
        &=&     G'.
\end{eqnarray}

Integrating now $(\ref{Gammavvv})$, using $(\ref{G'})$,
we obtain
\[
|\log \Omega^2(u,v)|\le |\log \Omega^2(u,0)|+G'V,
\]
and thus, since $|\log \Omega^2(u,0)|\le C$ for
some $C$, we have,
\begin{equation}
\label{cD}
0<c\le \Omega^2(u,v) \le D.
\end{equation}

Now, we turn to estimate the projection 
to the $p^v$-axis of the support of $f$.
We proceed by considering the geodesic equation.
Let $\gamma(s)$ be a geodesic crossing
$\{0\}\times[0,V]
\cup[0,U]\times\{0\}$ at $s=0$,
such that $\gamma'(0)$ is in the support
of $f$.
Let $p^v(s)$ denote the $\frac\partial{\partial v}$ 
component of the tangent vector of $\gamma$.
We have
\begin{equation}
\label{geodesic}
(p^v)'(s)=-\Gamma^{v}_{vv}(p^v)^2-\Gamma^{v}_{AB}p^Ap^B.
\end{equation}
using the Christoffel symbols in Appendix~\ref{Christoffels}.
Integrating $(\ref{geodesic})$,
we have now by $(\ref{GammaABv})$
\begin{eqnarray*}
p^v(s)  &=&     p^v(0)e^{-\int_0^s\Gamma^{v}_{vv}(p^v)d\tilde{s}}
                        -\int_0^s\Gamma^{v}_{AB}
                        p^A(\tilde{s})p^B(\tilde{s})
                        e^{-\int_{\tilde{s}}^s
                        \Gamma^{v}_{vv}(p^v)d\bar{s}}d\tilde{s}\\
        &=&     p^v(0)e^{-\int_{v(0)}^{v(s)}\Gamma^{v}_{vv}dv}
                        -\int_0^s\Gamma^{v}_{AB}
                        p^Ap^Be^{-\int_{v(\tilde{s})}^{v(s)}
                        \Gamma^{v}_{vv}dv}d\tilde{s}\\
        &=&     p^v(0)e^{-\int_{v(0)}^{v(s)}\Gamma^{v}_{vv}dv}
                        +\int_{v(0)}^{v(s)}
                        2(-\nu)\Omega^{-2} r\gamma_{AB}
                        p^Ap^Be^{-\int_{v(\tilde{s})}^{v(s)}
                        \Gamma^{v}_{vv}dv}(p^v)^{-1}dv.
\end{eqnarray*}
Thus, for $s'<s$, (replacing $0$ with $s'$)  we have,
by $(\ref{cD})$ and $(\ref{nubound})$, the inequality 
\begin{equation}
\label{ektimnsn2}
p^v(s) \le      p^v(s')e^{-\int_{v(s')}^{v(s)}\Gamma^{v}_{vv}dv}
                        +\int_{v(s')}^{v(s)}
                        2\tilde{N}c^{-1}r\gamma_{AB}
                        p^Ap^Be^{-\int_{v(\tilde{s})}^{v(s)}
                        \Gamma^{v}_{vv}dv} (p^v)^{-1}dv.
\end{equation}
Suppose $p^v(s)> 2\Sigma$ for some $0\le v(s)\le V$,
and let $s'$ be the last previous time $s>s'>0$
such that $ p^v(s')\ge 2\Sigma$, i.e., we have
$p^v(s^*)\ge 2\Sigma$ on $[s^*,s]$.
By $(\ref{ektimnsn2})$, $(\ref{G'})$, the angular momentum bound
$(\ref{am})$, and $(\ref{Gammavvv})$,
we have
\[
p^v(s) \le      2\Sigma e^{VG'}+2RX^2\tilde{N}c^{-1}
                        e^{VG'}V(2\Sigma)^{-1},
\]
i.e.
\begin{equation}
\label{ektimnsn}
p^v(s)\le \tilde{C}.
\end{equation}

We can now easily estimate
$T_{uu}$ pointwise:
\begin{eqnarray*}
T_{uu}  &=&     \int_0^\infty\int_{|\gamma_{AB}p^Ap^B|\le X}
                        r^2(p_u)^2f\frac{dp^v}{p^v}\sqrt{\gamma}dp^Adp^B\\
        &=&     (g_{uv})^2\int_0^\infty\int_{|\gamma_{AB}p^Ap^B|\le X}
                        r^2(p^v)^2f\frac{dp^v}{p^v}\sqrt{\gamma}dp^Adp^B\\
        &=&     4\nu^2\kappa^2\int_0^\infty\int_{|\gamma_{AB}p^Ap^B|\le X}
                        r^2(p^v)^2f\frac{dp^v}{p^v}\sqrt{\gamma}dp^Adp^B\\
        &=&     4\nu^2F\kappa^2\int_0^{\tilde{C}}
                        \int_{|\gamma_{AB}p^Ap^B|\le X}
                        r^2(p^v)dp^v\sqrt{\gamma}dp^Adp^B\\
        &\le&   16\pi R^2\nu^2F\tilde{C}^2X^2=\nu^2E\\
        &\le&   \tilde{N}^2E,
\end{eqnarray*}
in view of $(\ref{ektimnsn})$, $(\ref{nubound})$, $(\ref{kappa})$, $(\ref{r})$ and 
the angular momentum bound $(\ref{am})$. 
(Note that $T_{uu}\nu^{-2}\le E$ is a coordinate
invariant\footnote{i.e.~it does not depend on the normalization
of $u$} bound.)
Integrating $(\ref{puk})$, we obtain now
\[
\kappa\ge e^{-\int{4\pi r\frac{T_{uu}}{\nu^2}\nu du}}
\ge e^{-4\pi RE\tilde{N}U}.
\]
(Actually, we have in fact already estimated $\kappa$ from below
since $\kappa^{-1}=4(-\nu)\Omega^{-2}$.)

{}From the inequality
\[
p_up_v\le \frac12(p_u^2+p_v^2),
\]
we have
\[
T_{uv}\le\frac 12\left(T_{uu}+T_{vv}\right).
\]
This allows us to estimate $\partial_u\log\Omega^2=\Gamma_{uu}^u$:
\begin{eqnarray*}
|\Gamma_{uu}^u| &\le&   G+\left|\int{8\pi T_{uv}dv}-
\int{8\pi\kappa\nu r^{-2}\gamma^{AB}T_{AB}dv}\right|
                -\int{4\kappa mr^{-3}\nu dv}\\
        &\le&   G+4\pi \tilde{N}^2EV+\int{4\pi T_{vv}dv}
                -\int{4\kappa mr^{-3}\nu dv}\\
        &\le&   \bar{C}
\end{eqnarray*}
We can easily obtain an estimate now for $T_{vv}$. $\lambda$ can be bounded
by integrating (\ref{evol1}).


\section{$C^2$ estimates for the metric}
\label{c2sec}
In this section, we derive $C^2$ estimates on the metric and $C^1$
estimates for $f$.
The ideas of this section originate in~\cite{rr}.

It has already been shown that the following quantities are bounded:
$r$, $r^{-1}$, $m$, $m^{-1}$, $\kappa$, $\kappa^{-1}$, $\nu$, $\nu^{-1}$, 
$\lambda$, $\Omega$, $\Omega^{-1}$, all first order derivatives of $\Omega$,
all components of the energy-momentum tensor, and all Christoffel symbols in 
$(\ref{GammaABu})$--$(\ref{Gammavvv})$.
{}From these estimates and $(\ref{vu-e3is})$ and $(\ref{lambda-e3is})$,
it follows that $\partial_v\nu$ 
and $\partial_u\lambda$ are bounded, 
from $(\ref{pum})$ and $(\ref{pvm})$ it follows that $\partial_um$ and
 $\partial_vm$ are bounded,
and from $(\ref{kampulotnta})$, it follows that
$\partial_u\partial_v\Omega$ is bounded. 
Writing $\nu=-\frac14\Omega^2\kappa^{-1}$ and differentiating in $u$, 
we see from $(\ref{puk})$ that
$\partial_u\nu$ is bounded, while
writing $\kappa=-\frac14\Omega^2\nu^{-1}$ and differentiating in $v$, 
we see that
$\partial_v\kappa$ is bounded, and thus, from $(\ref{lambda=})$,
we see that $\partial_v\lambda$ is bounded.
These estimates 
and the formulas $(\ref{GammaABu})$--$(\ref{Gammavvv})$ 
allow us to control all first 
order derivatives of the Christoffel symbols,
except $\partial_u\Gamma^u_{uu}$ 
and $\partial_v\Gamma^v_{vv}$.

Since the components of the curvature tensor can be expressed in terms of
those derivatives of the Christoffel symbols which have already been 
estimated, we obtain bounds for all components of the curvature tensor in our
coordinates. The above estimates allow us to estimate the first derivatives 
of the exponential map on the tangent bundle. This, in turn allows one to 
estimate the derivatives of $f$ in terms of initial data.

We can, however,  argue more directly as follows. Let us abbreviate the
Vlasov equation (\ref{vlasoveq}) by $X(f)=0$ where $X$ is the Vlasov 
operator written in these coordinates. Note that $p^v$ is to be thought of 
as expressed in terms of $p^u$ and $p^A$ via the mass shell condition 
(\ref{massshell}).
Define $f_1=\partial_u f-p^u\partial_u\log\Omega^2\partial_{p^u}f$. 
Differentiating the Vlasov equation with respect to $v$, $p^u$ and $p^A$ 
gives the following equations:
\begin{eqnarray}
X(\partial_v f)&=&-(\partial_v p^v)\partial_v f+
(\partial_u\partial_v\log\Omega^2(p^u)^2
+\partial_v(-2\Omega^{-2}r\lambda)\gamma_{AB}p^Ap^B)\partial_{p^u} f
\nonumber\\
&&\hbox{}+2(\partial_v(\nu r^{-1})p^u+\partial_v(\lambda r^{-1})p^v
+\lambda r^{-1}\partial_v p^v)p^A
\partial_{p^A} f    \\
\nonumber
X(\partial_{p^u} f)&=&-\partial_u f-(\partial_{p^u} p^v)\partial_v f  
+2\partial_u\log\Omega^2p^u\partial_{p^u} f\\
&&\hbox{}+2(\nu r^{-1}
+\lambda r^{-1}\partial_{p^u} p^v)p^A\partial_{p^A}f
\\
X(p^D\partial_{p^D} f)&=&-p^D(\partial_{p^D}p^v)\partial_v f
-4\Omega^{-2}r\lambda\gamma_{AB} p^Ap^B\partial_{p^u} f \\
&&\hbox{}
+2r^{-1}\lambda p^D\partial_{p^D}p^vp^A
\partial_{p^A} f.
\end{eqnarray}
Differentiating the Vlasov equation with respect to $u$ gives the following
equation for $f_1$:
\begin{eqnarray}
X(f_1)&=&-p^u\partial_u(\log\Omega^2)X(\partial_{p^u}f)
-\partial_u p^v\partial_v f 
\nonumber\\
&+&(-p^up^v\partial_u\partial_v\log\Omega^2
-\partial_u\log\Omega^2(\partial_u\log\Omega^2(p^u)^2
+2\Omega^{-2}r\lambda\gamma_{AB}p^Ap^B)
\nonumber\\
&-&2\partial_u(\Omega^{-2}r\lambda)\gamma_{AB}p^Ap^B)\partial_{p^u} f  
\nonumber\\
&+&2(\partial_u(\nu r^{-1})p^u+\partial_u(\lambda r^{-1})p^v
+\partial_u p^v\lambda r^{-1})p^A
\partial_{p^A} f.
\end{eqnarray}
The quantity $X(\partial_{p^u}f)$ can be substituted for by one of the 
previous equations and $\partial_u f$ may be eliminated from the equations 
in favour of $f_1$. The result is a linear system of equations for the 
evolution of $(f_1,\partial_v f,\partial_{p^u} f, p^A\partial_{p^A} f)$ along 
the characteristics of the Vlasov equation. The coefficients are known to be 
bounded and so we can conclude that $\partial_u f$, $\partial_v f$,
$\partial_{p^u}f$ and $p^A\partial_{p^A}f$ are also bounded. (Note that since
$p^u$ and $p^v$ are bounded the derivative with respect to $X$ is uniformly
equivalent to a derivative along the characteristic with respect to $u$ or 
$v$ as parameter.)

{}From this, we immediately estimate $\partial_u T_{ab}$ and
$\partial_v T_{ab}$ pointwise. We now estimate $\partial_u\Gamma_{uu}^{u}$ 
by differentiating $(\ref{kampulotnta})$ in $u$
and integrating in $v$, and similarly,
$\partial_v\Gamma_{vv}^v$ by differentiating
in $v$ and integrating in $u$. Note that $|\partial_{p^A}|_\gamma$ can also
be bounded. This can be seen by passing from polar to Cartesian coordinates
and noting that the resulting metric components are $C^2$. As a consequence 
$f$ is $C^1$.

\section{The Proof of Theorem~\ref{extension}}
\label{theproof}
Let $N/2$ denote the sup of the norm
defined in Proposition~\ref{local}, where
the $\sup$ is taken now in all of $\mathcal{D}$.
By the estimates of the previous section,
we have that $N/2<\infty$. Let
$\delta$ be the constant of Proposition~\ref{local}
corresponding to $N$. 
Consider the point
$(U-\delta/2,V-\delta/2)$. Translate the coordinates
so that this point is $(0,0)$.
Since
$\mathcal{Q}$ is by definition open, by continuity,
there exists a $\delta>\delta^*>\delta/2$ such that
\[
\{0\}\times[0,\delta^*]
\cup [0,\delta^*]\times \{0\}
\subset
\mathcal{Q}
\]
and the 
assumptions of Proposition~\ref{local}
hold on
$\{0\}\times[0,\delta^*]
\cup [0,\delta^*]\times \{0\}$,
with $N$ and $\delta^*$ as already defined.
It follows that there exists a unique solution of
in
\[
\mathcal{E}=[0,\delta^*]\times[0,\delta^*].
\] 
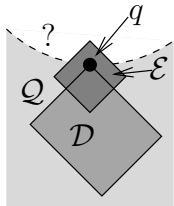
\[
\begin{picture}(0,0)%
\includegraphics{figure2.pstex}%
\end{picture}%
\setlength{\unitlength}{3947sp}%
\begingroup\makeatletter\ifx\SetFigFont\undefined%
\gdef\SetFigFont#1#2#3#4#5{%
  \reset@font\fontsize{#1}{#2pt}%
  \fontfamily{#3}\fontseries{#4}\fontshape{#5}%
  \selectfont}%
\fi\endgroup%
\begin{picture}(1066,1347)(4268,-4112)
\put(4351,-3436){\makebox(0,0)[lb]{\smash{\SetFigFont{12}{14.4}{\rmdefault}{\mddefault}{\updefault}{\color[rgb]{0,0,0}$\mathcal{Q}$}%
}}}
\put(5046,-2921){\makebox(0,0)[lb]{\smash{\SetFigFont{12}{14.4}{\rmdefault}{\mddefault}{\updefault}{\color[rgb]{0,0,0}$q$}%
}}}
\put(4671,-3721){\makebox(0,0)[lb]{\smash{\SetFigFont{12}{14.4}{\rmdefault}{\mddefault}{\updefault}{\color[rgb]{0,0,0}$\mathcal{D}$}%
}}}
\put(5176,-3301){\makebox(0,0)[lb]{\smash{\SetFigFont{12}{14.4}{\rmdefault}{\mddefault}{\updefault}{\color[rgb]{0,0,0}$\mathcal{E}$}%
}}}
\put(4496,-3091){\makebox(0,0)[lb]{\smash{\SetFigFont{12}{14.4}{\rmdefault}{\mddefault}{\updefault}{\color[rgb]{0,0,0}$?$}%
}}}
\end{picture}

\]
Thus the solution coincides in $\mathcal{E}\cap\mathcal{Q}$
by uniqueness. One sees that
$\mathcal{E}\cup\mathcal{Q}$ is clearly the quotient
of a development
of initial data. By maximality of $\mathcal{M}$, we must
have $\mathcal{E}\cup\mathcal{Q}\subset\mathcal{Q}$.
Thus, in particular, in the old coordinates we have
$(U,V)\in\mathcal{Q}$, and the theorem holds with $q=(U,V)$.

\section{Applications}
\label{appli}
We will say that a spherically symmetric maximal development
has a \emph{black hole},
if $\mathcal{I}^+$ is \emph{complete} in the sense of~\cite{chr:givp},\footnote{
See~\cite{md} for a definition of $\mathcal{I}^+$ in this context.} and
if $J^-(\mathcal{I}^+)$ has a non-empty complement.
 
We have shown that the results of~\cite{md} apply to our matter model.
In particular, the fact that the complement of $J^-(\mathcal{I}^+)$ is
non-empty implies the completeness of null infinity. That this set is 
non-empty can be inferred in turn from the existence
of a single trapped or marginally trapped surface. 
Asymptotically flat
spherically symmetric solutions of the Einstein-Vlasov system possesing a
trapped surface were constructed in~\cite{adr1}. Thus we have
\begin{corollary}
There exist solutions of the Einstein-Vlasov system which develop from 
regular initial data and contain black holes.
\end{corollary}

The fundamental open question in gravitational collapse is to show that
generically, either the solution is future geodesically complete or a black 
hole forms. In view of~\cite{md} and the results of this paper we have
\begin{corollary}
Suppose that for generic initial data, the maximal development either
contains a trapped surface or marginally trapped surface, or is
future causally geodesically complete. Then weak cosmic censorship is
true.
\end{corollary}

Thus, weak cosmic censorship can be reduced to a slightly weaker version
of Christodoulou's \emph{trapped surfaces conjecture}. As remarked in the 
Introduction, this suggests
a local approach to its proof (cf.~\cite{chr:ins}). 

\section{Acknowledgement} We gratefully acknowledge the support of the Erwin
Schr\"odinger Institute, Vienna, where an important part of this research
was carried out.

\appendix
\section{The Christoffel symbols}
\label{Christoffels}
Note:
\[
g_{uv}=-\frac12\Omega^2,
\]
\[
g^{uv}=-2\Omega^{-2},
\]
\[
\Omega^{2}=-4\kappa\nu.
\]
The nonvanishing Christoffel symbols are given by:
\begin{equation}
\label{GammaABu}
\Gamma_{AB}^u=-g^{uv}r\lambda\gamma_{AB},
\end{equation}
\begin{equation}
\label{GammaABv}
\Gamma_{AB}^v=-g^{uv}r\nu\gamma_{AB},
\end{equation}
\begin{equation}
\label{GammaAvA}
\Gamma_{Bv}^A=\lambda r^{-1}\delta^A_B,
\end{equation}
\begin{equation}
\label{GammaAuA}
\Gamma_{Bu}^A=\nu r^{-1}\delta^A_B,
\end{equation}
\begin{equation}
\label{Gammauuu}
\Gamma_{uu}^u=\partial_u\log \Omega^2,
\end{equation}
\begin{equation}
\label{Gammavvv}
\Gamma_{vv}^v=\partial_v\log \Omega^2.
\end{equation}
In fact the Christoffel symbols $\Gamma_{AB}^C$, which depend on a choice
of coordinates on the spheres of symmetry need not vanish but the 
expressions for them are not needed in this paper.


\section{Proof of Proposition~\ref{local}}
\label{localproof}
The proof of local existence follows from simpler considerations
than the proof of the estimates of Sections~\ref{enestsec}--\ref{c2sec}.
In particular, one does not need to consider energy estimates,
for one can recover naive pointwise estimates using the smallness parameter.
As in Section \ref{c2sec}, the idea of~\cite{rr}
again makes its appearance, to show $C^1$
bounds on $f$ directly from $C^0$ bounds on the curvature, before bounding the
$C^2$ norm of the metric.
Since all these methods have appeared before, we will only sketch the details
here.

Let initial data be fixed. 
Define the space 
\[
A\subset C^2([0,\delta]\times[0,\delta])\times C^1([0,\delta]\times[0,\delta]),
\]
for $\delta$ to be determined later,
consisting of
all twice continuously differentiable nonnegative functions $r$,
continuously differentiable nonnegative functions $\Omega$, 
extending the
prescribed values, such that 
\begin{equation}
\label{rupdown}
N^{-1}/2 \le r\le 2N,
\end{equation}
\begin{equation}
\label{Nupdown}
N^{-1}/2 \le \Omega \le 2N,
\end{equation}
\begin{equation}
\label{tapolla1}
\sup\{
|\partial_ur|,
|\partial_vr|, 
|\partial^2_{u}r|,
|\partial^2_{v}r|\}\le 2N
\end{equation}
\begin{equation}
\label{tapolla2}
\sup\{
|\partial_u\Omega|,
|\partial_v\Omega|\}\le 2N.
\end{equation}

Consider the subset $B\subset A$, consisting of those 
$(r,\Omega)$ for which $\Omega$ is $C^2$, and 
for which
\begin{equation}
\label{2ndderivs}
\sup\{
|\partial_u^2\Omega|,
|\partial_v^2\Omega|,
|\partial_u\partial_v\Omega|\}\le 2N.
\end{equation}

Note that the closure of 
$B$ in $A$, denoted $\overline{B}$, consists
of $(r,\Omega)$ such that $\partial_u\Omega$,
$\partial_v\Omega$, are Lipschitz, with Lipschitz constants
given by the above.

We shall define in the next few paragraphs a continuous map 
$\Phi:\overline{B}\to A$ 
taking $(r, \Omega)$ to $(\tilde r,
\tilde{\Omega})$.

Given $r$, $\Omega$, first, let $f$ be defined to 
solve the Vlasov equations on the metric defined
by $r$ and $\Omega$, with given initial conditions.
Note that since the Christoffel symbols of this metric are Lipschitz,
it follows that geodesics can be defined, and thus $f$ can be defined
by the requirement that it is preserved by geodesic motion. 
It follows immediately that
\begin{equation}
\label{fbound}
0\le f\le N,
\end{equation}
and, after appropriately restricting to sufficiently small $\delta$,
it follows easily by integration of the geodesic equations that
\begin{equation}
\label{sbound}
S\le 2N.
\end{equation}
In the case 
where $(r,\Omega)\in B$, we have that $f$ is in fact $C^1$,
since the exponential map is differentiable. 
If $\delta$ is chosen sufficiently small, it is clear 
from $(\ref{rupdown})$--$(\ref{2ndderivs})$ that,
in this case,
we can arrange for
\begin{equation}
\label{fderivbound}
\sup\{|\partial_vf|,
|\partial_uf|,
|\partial_{p_u}f|,
|\partial_{p^A}f|_\gamma\}
\le 2N.
\end{equation}

Given now $f$, we can define $T^{uv}$, $T^{vv}$, $T^{uu}$
in the standard way. 
In view of $(\ref{rupdown})$--$(\ref{tapolla2})$,
$(\ref{fbound})$, and $(\ref{sbound})$, 
these terms can be estimated.
Now, set 
$\nu=\partial_ur$, $\lambda=\partial_vr$.
We define $\tilde{r}$ by
\begin{equation}
\label{tildenu}
\tilde{r}(u,v)=r(u,0)+r(0,v)-r(0,0)+
\int_0^u\int_0^v-\frac14 r^{-2}\Omega^2-\frac1r\lambda\nu
+4\pi r\Omega^4 T^{uv}dudv
\end{equation}

By appropriate differentiation of $(\ref{tildenu})$, it is clear 
from our bounds thus far that  
we can define and estimate $\tilde\nu=
\partial_u \tilde{r}$, $\tilde\lambda=\partial_v \tilde{r}$,
and $\partial_u\partial_v \tilde{r}$.
We can retrieve the bound $(\ref{rupdown})$ for $\tilde{r}$
by integration of the $\tilde\nu$, after restricting to small $\delta$.
For $(r,\Omega)\in B\subset\bar{B}$, it is clear we can also
define and estimate
$\partial_u^2 \tilde r$,
$\partial_v^2 \tilde r$, by
differentiating $(\ref{tildenu})$ twice in $u$ or twice in $v$, 
in view of the fact that
all other derivatives, including $\partial_u T^{uv}$, $\partial_u\nu$, etc.,
are clearly defined and bounded, in view of $(\ref{fderivbound})$,
and since these derivatives are defined initially.
By appropriate choice of $\delta$, we can clearly arrange--for
$(r,\Omega)\in B$--so as
to retrieve the bound $(\ref{tapolla1})$.

Define now ${\tilde\Omega}>0$ by the relation 
\begin{eqnarray}
\label{Omega}
\log {\tilde\Omega}^2&=&\log\Omega^2(u,0)+\log \Omega^2(0,v)-\log\Omega^2(0,0)
\\
\nonumber
&&\hbox{}+
\int_0^u\int_0^v{(-8\pi T_{uv}+\frac12\Omega^2\tilde r^{-2}+2\tilde{r}^{-2}
\tilde\lambda\tilde \nu-2\pi\Omega^2
\tilde{r}^{-2}\gamma^{AB}T_{AB})dudv}.
\end{eqnarray}
Again, for small enough $\delta$, it is clear that one can arrange
for $\tilde{\Omega}$ to satisfy $(\ref{Nupdown})$. 

Differentiating $(\ref{Omega})$ appropriately, in view of the initial
conditions for $\tilde\Omega$, it follows
that, for $(r,\Omega)\in \overline{B}$, 
$\tilde\Omega$ is $C^1$, and for $\delta$ small enough
satsfies $(\ref{tapolla2})$,
while 
for $(r,\Omega)\in B$, $\tilde\Omega$ is $C^2$, and
for $\delta$ small enough, satisfies $(\ref{2ndderivs})$.

Thus, we have shown that
after judicious choice of $\delta$, $\Phi$ maps $B$ to itself.
By continuity, it maps $\overline{B}$ to itself.

The map $\Phi$ can easily be shown to be a contraction in $B$
for the norm of $A$, i.e., we can show that
\begin{equation}
\label{contraction}
d_A((\tilde{r}_1,\tilde{\Omega}_1),(\tilde{r}_2,\tilde{\Omega}_2))\le 
\epsilon d_A((r_1,\Omega_1),(r_2,\Omega_2)),
\end{equation}
for an $\epsilon<1$ and all $(r_i,\Omega_i)\in B$.
To see this, define first $f_i$, corresponding to $(r_i,\Omega_i)$. 
Let $\Gamma_i$ denote an arbitrary Christoffel symbol for $(r_i, \Omega_i)$. 
We clearly have
\[
|\Gamma_1-\Gamma_2|\le Cd_A ((r_1,\Omega_1),(r_2,\Omega_2)).
\]
We easily obtain
\[
|f_1-f_2|\le 
C\delta \sup_{\Gamma}|\Gamma_1-\Gamma_2|\sup_{i=1,2}(|\partial f_i|+|f_i|).
\]
Clearly we can also bound
$\sup|T_1^{uv}-T_2^{uv}|\le C \sup|f_1-f_2|$.
One bounds $(\nu_1-\nu_2)$ by expressing 
$\partial_v(\tilde{\nu}_1-\tilde{\nu}_2)$ as a linear combination of
$\Omega_1-\Omega_2$, $r_1-r_2$, $\nu_1-\nu_2$, $\lambda_1-\lambda_2$
and $(T_1^{uv}-T_2^{uv})$ with bounded coefficients.
One immediately obtains a similar bound for
$\sup|\tilde{r}_1-\tilde{r}_2|$.
The terms 
$\sup|\partial_u\tilde{r}_1-\partial_u\tilde{r}_2|$,
$\sup|\partial_v\tilde{r}_1-\partial_v\tilde{r}_2|$,
and $\sup|\partial_u\partial_v \tilde{r}_1-\partial_u\partial_v\tilde{r}_2|$,
can be handled in the same way.
One then obtains a bound of the above form for 
$\sup|\partial_v\log {\tilde \Omega}_1^2-\partial_v\log {\tilde \Omega}_2^2|$,
and similarly for $\sup|\partial_u\log {\tilde \Omega}_1^2-
\partial_u\log {\tilde \Omega}_2^2|$. 
Either of these bounds of course implies
a bound for $\sup|{\tilde \Omega}_1^2-{\tilde \Omega}_2^2|$.

To bound $\sup|\partial^2_{u}\tilde{r}_1-\partial^2_{u}\tilde{r}_2|$,
we compute
\begin{eqnarray}
\label{megalo}
\nonumber
\partial^2_u\tilde{r}&=&\partial^2_u\tilde{r}|_{v=0}+\int_0^v
\partial_u\left(-\frac14r^{-2}\Omega^2-r^{-1}\lambda\nu\right)
\\
&&\hbox{}+4\pi\partial_u(r\Omega^4)T^{uv}
+4\pi r\Omega^4\partial_uT^{uv}dv\\
\nonumber
&=&
\partial^2_u\tilde{r}|_{v=0}+\int_0^v
\partial_u\left(-\frac14r^{-2}\Omega^2-r^{-1}\lambda\nu\right)
+4\pi\partial_u(r\Omega^4)T^{uv}\\
\nonumber
&&\hbox{}-4\pi r\Omega^4\partial_vT^{vv}
+4\pi r\Omega^4(\sum T\cdot \Gamma)dv\\
\nonumber
&=&
\partial^2_u\tilde{r}|_{v=0}-4\pi r\Omega^4T^{vv}(u,v)+4\pi r
\Omega^4T^{uv}(u,0)\\
\nonumber
&&\hbox{}+\int_0^v
\partial_u\left(-\frac14r^{-2}\Omega^2-r^{-1}\lambda\nu\right)
+4\pi\partial_u(r\Omega^4)T^{uv}\\
&&\hbox{}
+4\pi\partial_v(r\Omega^4)T^{vv}
+4\pi r\Omega^4(\sum T\cdot \Gamma)dv.
\end{eqnarray}
Here we have used the equation $\nabla_a T^{ab}=0$,
which follows from the Vlasov equation, and we have integrated by parts.
It is now clear that estimates for differences follow as before.
We argue in an entirely analogous way for
$\sup|\partial^2_v\tilde{r}_1-\partial^2_v\tilde{r}_2|$.

After restricting to sufficiently small $\delta$, all constants
in the above bounds can be made small. We
thus have indeed shown $(\ref{contraction})$.
It follows by continuity that $\Phi$ is also a contraction on
$\overline{B}\subset A$, and thus, since $\overline{B}$ is
closed, has a fixed point in $\overline{B}$. 

Given such a fixed point $(r, \Omega)$, define $f$ as before.
To show that $(r, \Omega, f)$ corresponds to a solution of the equations,
we have basically only to show that $f$ and
$\partial_u\Omega$, $\partial_v\Omega$,
which \emph{a priori} are Lipschitz,
are in fact $C^1$. (In particular, from this it will follow that the
constraint equations $(\ref{constr1})$--$(\ref{constr2})$
are also satisfied.)
But, in view of the fact that $f$ is initially $C^1$,
it follows that $f$ is $C^1$ if the exponential map is $C^1$.
(The $C^2$ compatibility condition is used at the point.)
But this latter fact follows from the continuity of the curvature, 
as shown in Exercise 6.2 of Chapter V of \cite{hartman}
\footnote{If the
reader does want to apply to this fact, then one can argue as
follows: in view of the computations above, in the space $B$, we have
that curvature is in fact $C^1$ with estimates; since derivatives
of the exponential map are computed by integrating curvature
on geodesics, and geodesics certainly depend $C^1$ on their initial
conditions, in view of the fact that the Christoffel symbols
are $C^1$ with bounds in $B$, 
it follows that we have $C^2$ estimates for
the exponential map in $B$, and thus by an easy compactness argument,
the exponential map of the fixed point must be $C^1$. There is only
one catch with this argument: $r$ and $\Omega^2$ have to be
assumed to be initially $C^3$ to differentiate $(\ref{Omega})$ and
$(\ref{megalo})$ three times.}. That the curvature is continuous  
follows by computation, since $r$ is $C^2$, $\Omega$ is $C^1$ 
and $\partial_u\partial_v\Omega$
is $C^0$, and $\partial^2_u\Omega$ and $\partial^2_v\Omega$
do not appear in the expressions for curvature.
{}From the $C^1$ property of $f$, the $C^2$ property of
$\Omega$ follows immediately.
Similarly, higher regularity follows immediately if
it is assumed. 
$\Box$

\end{document}